\documentclass[reqno,draft,10pt]{amsart}

\usepackage{verbatim}

\theoremstyle{plain}
\newtheorem{lemma}{Lemma}
\newtheorem{theorem}{Theorem}
\newtheorem{corollary}{Corollary}

\newtheorem{proposition}{Proposition}
\newtheorem{definition}{Definition}

\theoremstyle{remark}

\renewcommand{\labelenumi}{(\theenumi)}

\newcommand*  {\R} {{\mathbb R}}

\newcommand*  {\C} {{\mathbb C}}

\newcommand*{\norm}[3][{\vphantom 1}]{\lVert #2 \rVert_{#3}^{#1}}

\newcommand*{\abs}[1]{\lvert #1 \rvert}
\newcommand*{\babs}[1]{\bigg \lvert #1 \bigg \rvert}

\newcommand*{\ang}[1]{\left\langle #1 \right\rangle}

\newcommand{\nm}   [1]{\norm{#1}{}{}}
\newcommand{\ddt}     {\frac{d}{dt}}
\newcommand{\pdt}  [1]{\frac{d #1}{d t}}
\newcommand{\half}    {\frac{1}{2}}

\begin{document}

\title[A Note on the Regularity of Inviscid Shell Model of Turbulence]%
{A Note on the Regularity of Inviscid Shell Model of Turbulence}

\date{July 6, 2006}

\author[P. Constantin]{Peter Constantin}
\address[P. Constantin]%
{Department of Mathematics \\
 The University of Chicago \\
 Chicago, IL 60637 \\
 USA}
\email{const@math.uchicago.edu}

\author[B. Levant]{Boris Levant}
\address[B. Levant]%
{Department of Computer Science and Applied Mathematics \\
 Weizmann Institute of Science \\
 Rehovot, 76100 \\
 Israel}
\email{boris.levant@weizmann.ac.il}

\author[E. S. Titi]{Edriss S. Titi}
\address[E. S. Titi]%
{Department of Mathematics and
 Department of Mechanical and Aerospace Engineering \\
 University of California \\
 Irvine, CA 92697 \\
 USA \\
 Also, Department of Computer Science and Applied Mathematics \\
 Weizmann Institute of Science \\
 Rehovot, 76100 \\
 Israel}
\email{etiti@math.uci.edu and edriss.titi@weizmann.ac.il}


\begin{abstract}
In this paper we continue the analytical study of the sabra shell
model of energy turbulent cascade initiated in \cite{CLT05}. We
prove the global existence of weak solutions of the inviscid sabra
shell model, and show that these solutions are unique for some short
interval of time. In addition, we prove that the solutions conserve
the energy, provided that the components of the solution satisfy
$\abs{u_n} \le C k_n^{-1/3} (\sqrt{n} \log(n+1))^{-1}$, for some
positive absolute constant $C$, which is the analogue of the
Onsager's conjecture for the Euler's equations. Moreover, we give a
Beal-Kato-Majda type criterion for the blow-up of solutions of the
inviscid sabra shell model and show the global regularity of the
solutions in the ``two-dimensional'' parameters regime.
\end{abstract}

\maketitle

\section{Introduction}

Shell models of turbulence have attracted  interest as useful
phenomenological models that retain certain features of the
Navier-Stokes and Euler equations. Their central computational
advantage is the parameterization of the fluctuation of a
turbulent field in an octave of wave numbers $\lambda^{n} <
\abs{k_n} \le \lambda^{n+1}$ by very few representative variables.
This range of wave numbers is called a shell and the variables
retained are called shell variables. Like in the Fourier
representation of Navier-Stokes equations (NSE), the time
evolution of the shell variables is governed by an infinite system
of coupled ordinary differential equations with quadratic
nonlinearities, with forcing applied to the large scales and
viscous dissipation effecting the smaller ones. Because of the
very reduced number of interactions in each octave of wave
numbers, the shell models are drastic modification of the original
NSE in  Fourier space.

The main objective of this work is to investigate the question of
existence, uniqueness and regularity of solutions of the inviscid
sabra shell model of turbulence. This model was introduced in
\cite{LP98} and its viscous version was studied analytically in
\cite{CLT05}. It is worth noting that the results of this article
apply equally well to the well-known Gledzer-Okhitani-Yamada (GOY)
shell model, introduced in \cite{OY89}. For other shell models see,
e.g., \cite{BJPV98}, \cite{Gl73}, \cite{Fr95}, \cite{OY89}. A recent
review of the subject emphasizing  the applications of the shell
models to the study of the energy-cascade mechanism in turbulence
can be found in \cite{Bi03}.

The sabra shell model of turbulence describes the evolution of
complex Fourier-like components of a scalar velocity field denoted
by $u_n$. The associated one-dimensional wavenumbers are denoted by
$k_n$, where the discrete index $n$ is referred to as the ``shell
index''. The equations of motion of the viscous sabra shell model of
turbulence have the following form
\begin{equation} \label{eq_sabra}
\frac{d u_n}{d t} = i (a k_{n+1} u_{n+2} u_{n+1}^* + b k_{n}
u_{n+1} u_{n-1}^* - c k_{n-1} u_{n-1} u_{n-2}) - \nu k_n^2 u_n +
f_n,
\end{equation}
for $n = 1, 2, 3, \dots$, and the boundary conditions are $u_{-1}
= u_{0} = 0$. The wave numbers $k_n$ are taken to be
\begin{equation} \label{eq_freq}
k_n = k_0 \lambda^n,
\end{equation}
with $\lambda  > 1$ being the shell spacing parameter, and $k_0
> 0$. Although the equation does not capture any geometry, we will
consider $L = k_0^{-1}$ as a fixed typical length scale of the
model. In an analogy to the NSE $\nu > 0$ represents a kinematic
viscosity and $f_n$ are the Fourier components of the forcing.

The choice of the non-linear term in the equation of the sabra model
(\ref{eq_sabra}) which contains only the local interaction between
the shells, can be justified in the context of the Kolmogorov theory
of homogeneous turbulence (see \cite{Gal02_Thebook}, \cite{Fr95}).
The theory states that there is no interchange of energy between the
modes of the velocity field with wavenumbers separated at least by
``an order of magnitude''.

The three parameters of the model $a, b$ and $c$ are real. In
order for the sabra shell model to be a system of the hydrodynamic
type we require that in the inviscid ($\nu = 0$) and unforced
($f_n = 0$, $n = 1, 2, 3, \dots$) case the model will have at
least one quadratic invariant. Requiring conservation of the
energy
\begin{equation} \label{eq_energy}
\mathrm{E} = \sum_{n=1}^\infty \abs{u_n}^2
\end{equation}
leads to the following relation between the parameters of the
model, which we will refer to it as the energy conservation
condition
\begin{equation} \label{eq_energy_cons_assumption}
a + b + c = 0.
\end{equation}
Moreover, in the inviscid and unforced case the model possesses
(formally) another quadratic invariant
\begin{equation} \label{eq_helicity}
\mathrm{W} = \sum_{n=1}^\infty \bigg( \frac{a}{c} \bigg)^n
\abs{u_n}^2.
\end{equation}

The ``physically'' relevant range of parameters is $\abs{a/c}
> 1$ (see \cite{LP98} for details). For $-1 < \frac{c}{a} < 0$
the quantity $\mathrm{W}$ is not sign-definite and therefore it is
common to associate it with the helicity -- in an analogy to the
$3$D turbulence. In that regime we can rewrite the relation
(\ref{eq_helicity}) in the form
\begin{equation} \label{eq_helicity2}
\mathrm{W} = \sum_{n=1}^\infty (-1)^n k_n^\alpha \abs{u_n}^2,
\end{equation}
for
\begin{equation}\label{eq_alpha}
\alpha = \log_\lambda \babs{\frac{a}{c}}.
\end{equation}
We call parameters regime corresponding to $0 < \frac{c}{a} < 1$ the
$2$D regime. This is because in that case the second conserved
quadratic quantity $\mathrm{W}$ is non-negative and can be
identified with the enstrophy in $2$D turbulent flows. We can
rewrite the expression (\ref{eq_helicity}) in the form
\begin{equation} \label{eq_helicity2_new}
\mathrm{W} = \sum_{n=1}^\infty k_n^\alpha \abs{u_n}^2,
\end{equation}
where $\alpha$ is also defined by the equation (\ref{eq_alpha}).

For the parameters satisfying $\frac{a}{c} = -\lambda$ the sabra
shell model becomes ``purely three-dimensional'', where the quantity
(\ref{eq_helicity}) scales like the helicity in the $3$D
Navier-Stokes turbulence. It was found (see \cite{DiMo96}) that in
that case the energy spectrum in the inertial range of the GOY shell
model (this is also true for the sabra model) has the traditional
Kolmogorov scaling law $k_n^{-5/3}$. Moreover, while the parameters
of the model satisfy $\frac{a}{c} = \lambda^2$, the quantity
(\ref{eq_helicity2}) scales like the enstrophy in the Navier-Stokes
$2$D turbulence. Therefore, this parameters values are usually
referred as the ``purely two-dimensional'' regime. In that case the
energy spectrum of the sabra shell model (see \cite{GLP02},
\cite{DiMo96}) obeys the scaling law $k_n^{-3}$, which is exactly
the Kraichnan's law of the $2$D developed turbulence.

The famous question of global well-posedness of the $3$D
Navier-Stokes and Euler equations is a major open problem. In
\cite{CLT05} we showed global regularity of weak and strong
solutions of (\ref{eq_sabra}) and smooth dependence on the initial
data for the case $\nu > 0$. In this work we address the question of
existence of regular solutions of the inviscid ($\nu = 0$) sabra
shell model (\ref{eq_sabra}). First, we prove the global in time
existence of weak solutions with finite energy. Similar results for
the inviscid GOY shell model were obtained recently in
(\cite{BBBF06_Stochastic}). The existence of weak solutions for
inviscid hydrodynamic equations with only energy conservation is not
known. The only other analogue of the $3$D Euler equations known to
posses weak solutions of such type is the inviscid surface
quasi-geostrophic equation (see \cite{CCW01_Quasi} and
\cite{Resn95_Quasi}). Next, we show that every weak solution $u(t) (u_1(t), u_2(t), \dots)$ conserve the energy provided that the
components of the solution satisfy the decay estimate
\[
\abs{u_n} \le C k_n^{-1/3} (\sqrt{n} \log(n+1))^{-1},
\]
for some positive absolute constant $C$, namely, provided it is
regular enough. A similar result for the solutions of Euler
equations is known as the Onsager's conjecture (see
\cite{Onsa49_Hydro}) and it was proved in \cite{CET94_Onsager} (see
also \cite{DuchRob00_Onsager}, \cite{Eyink94}). We also give the
criteria for the weak solutions to remain unique in certain
regularity class.

Next, we show that if the initial data is sufficiently smooth, then
the weak solutions are smooth and unique for a short period of time.
Similar results were obtained in the context of other discrete
models of Euler equations (see, e.g., \cite{FriPav04},
\cite{KaPav05}, \cite{KiZla05_Blowup}, \cite{Wal05}). The well known
Beale-Kato-Majda theorem (see \cite{BKM84}, \cite{MB02}) gives a
criterion for the blow-up of the initially smooth solutions of the
$3$D Euler equations. In section~\ref{sec_bkm} we establish a
similar criterion for the inviscid sabra shell model equations and
use it to show the global regularity of the solutions of the model
in the $2$D parameters regime. This picture is consistent with what
is known about the global regularity of solutions of the $2$D Euler
equations (see, e.g., \cite{Bard72_Euler}, \cite{Bard72_Euler2},
\cite{MB02}, \cite{MP94}, \cite{Yu63}, and references therein).

Analytic and numerical study of loss of regularity of solutions of
the inviscid sabra shell model of turbulence, as well as the
dissipation anomaly phenomena in that model, is the subject of
ongoing research (\cite{CLT06_Dissipation}).

\section{Preliminaries and Functional Setting}

We repeat here for the sake of self consistency the functional
settings introduced in section $2$ of~\cite{CLT05}. In particular,
the Proposition~\ref{prop_non_linear} is a slightly more generalized
version of the Proposition $1$ of~\cite{CLT05}.

Following the classical treatment of the NSE and Euler equations,
and in order to simplify the notation we are going to write the
system (\ref{eq_sabra}) in the following functional form
\begin{subequations}
\label{eq_abstract_model}
\begin{gather}
\pdt{u} + \nu A u + B(u, u) = f \label{eq_shell_model} \\
u(0) = u^{in}, \label{eq_initial_cond}
\end{gather}
\end{subequations}
in a Hilbert space $H$. The linear operator $A$ as well as the
bilinear operator $B$ will be defined below. In our case, the
space $H$ will be the sequences space $\ell^2$ over the field of
complex numbers $\C$. For every $u, v\in H$, the scalar product
$(\cdot, \cdot)$ and the corresponding norm $\abs{\cdot}$ are
defined as
\[
(u, v) = \sum_{n=1}^\infty u_n v_n^*, \;\;\; \abs{u} = \bigg(
\sum_{n=1}^\infty \abs{u_n}^2 \bigg)^{1/2}.
\]
We denote by $\{\phi_j\}_{j=1}^\infty$ the standard canonical
orthonormal basis of $H$, i.e. all the entries of $\phi_j$ are
zero except at the place $j$ it is equal to $1$.

The linear operator $A : D(A) \to H$ is a positive definite,
diagonal operator defined through its action on the elements of
the canonical basis of $H$ by
\[
A \phi_j = k_j^2 \phi_j,
\]
where the eigenvalues $k_j^2$ satisfy the equation (\ref{eq_freq}).
The space
\[
D(A) = \{ u\in H \;:\; \abs{A u}^2 = \sum_{n=1}^\infty k_n^4
\abs{u_n}^2 < \infty \},
\]
is the domain of $A$ and is a dense subset of $H$. Moreover, it is
a Hilbert space, when equipped with the graph norm
\[
\norm{u}{D(A)} = \abs{Au}, \;\;\; \forall u\in D(A).
\]
Using the fact that $A$ is a positive definite operator, we can
define the powers $A^s$ of $A$ for every $s\in \R$
\[
\forall u = (u_1, u_2, u_3, \dots), \;\;\; A^s u = (k_1^{2s} u_1,
k_2^{2s} u_2, k_3^{2s} u_3, \dots).
\]

Furthermore, we define the spaces
\begin{equation} \label{eq_vd_spaces}
V_s := D(A^{s/2}) = \{ u = (u_1, u_2, u_3, \dots) \;:\;
\sum_{j=1}^\infty k_j^{2s} \abs{u_j}^2 < \infty \},
\end{equation}
which are Hilbert spaces equipped with the scalar product
\[
(u, v)_{s} = (A^{s/2} u, A^{s/2} v), \;\;\; \forall u, v\in
D(A^{s/2}),
\]
and the norm $\abs{u}_s^2 = (u, u)_s$, for every $u\in
D(A^{s/2})$. Clearly
\[
V_{s} \subseteq V_0 = H \subseteq V_{-s}, \;\;\; \forall s > 0.
\]
Note that the dual space of $V_s$, for every $s\in \R$, is $V_s' V_{-s}$. We denote the action of the element in $u\in V_{-s}$ on
$v\in V_{s}$ by
\[
\ang{u, v}_s = (A^{-s/2} u, A^{s/2} v) = \sum_{n=1}^\infty u_n
v_n^*.
\]

The case of $s = 1$ is of a special interest for us. We denote $V D(A^{1/2})$ a Hilbert space equipped with a scalar product and norm
\[
((u, v)) = (A^{1/2}u, A^{1/2}v), \;\;\; \nm{u}^2 = ((u, v)),
\]
for every $u, v\in V$. The action of the element $u\in V_{-1} = V'$
on $v\in V$ is denoted by
\[
\ang{u, v} = (A^{-1/2}u, A^{1/2}v) = \sum_{n=1}^\infty u_n v_n^*.
\]

Before proceeding and defining the bilinear term $B$, let us
introduce the sequence analogue of Sobolev functional spaces.

\begin{definition} \label{def_sobolev}
For $1 \le p \le \infty$ and $m\in \R$ we define sequence spaces
\[
w^{m, p} := \{ u = (u_1, u_2, \dots) \;:\; \nm{A^{m/2} u}_p \bigg(
\sum_{n=1}^\infty (k_n^m \abs{u_n})^p \bigg)^{1/p} < \infty \},
\]
for $1 \le p < \infty$, and
\[
w^{m, \infty} := \{ u = (u_1, u_2, \dots) \;:\; \nm{A^{m/2}
u}_\infty = \sup_{1 \le n\le \infty} (k_n^m \abs{u_n}) < \infty \}.
\]
For $u\in w^{m, p}$ we define its norm
\[
\nm{u}_{w^{m, p}} = \nm{A^{m/2} u}_p,
\]
where $\nm{\cdot}_p$ is the usual norm in the $\ell^p$ sequence
space. The special case of $p = 2$ and $m\ge 0$ corresponds to the
sequence analogue of the classical Sobolev space, which we denote
by
\[
h^m = w^{m, 2}.
\]
Those spaces are Hilbert with respect to the norm defined above
and its corresponding inner product.
\end{definition}

The above definition immediately implies that $h^{d} = V^{d}$, for
all $d$. Moreover, $V_d \subset w^{d, \infty}$ and the inclusion map
is continuous because
\[
\nm{u}_{w^{d, \infty}} = \nm{A^{d/2} u}_\infty \le \nm{A^{d/2} u}_2
= \abs{u}_d.
\]

The bilinear operator $B(u, v)$ will be defined formally in the
following way. Let $u, v\in H$ be of the form $u \sum_{n=1}^\infty u_n \phi_n$ and $v = \sum_{n=1}^\infty v_n
\phi_n$. Then
\begin{align}
B(u, v) = - i \sum_{n=1}^\infty \bigg( a k_{n+1} v_{n+2} u_{n+1}^*
& + b k_{n} v_{n+1} u_{n-1}^* + \notag \\ & + a k_{n-1} u_{n-1}
v_{n-2} + b k_{n-1} v_{n-1} u_{n-2} \bigg) \phi_n,
\end{align}
where here again $u_0 = u_{-1} = v_0 = v_{-1} = 0$. It is easy to
see that our definition of $B(u, v)$, together with the energy
conservation condition (\ref{eq_energy_cons_assumption}) imply
that
\[
B(u, u) = - i \sum_{n=1}^\infty \bigg( a k_{n+1} u_{n+2} u_{n+1}^*
+ b k_{n} u_{n+1} u_{n-1}^* - c k_{n-1} u_{n-1} u_{n-2} \bigg)
\phi_n,
\]
which is consistent with (\ref{eq_sabra}). In~\cite{CLT05} we
showed that indeed our definition of $B(u, v)$ makes sense as an
element of $H$, whenever $u\in H$ and $v\in V$ or $u\in V$ and
$v\in H$. The next Proposition is a slightly generalized version
of Proposition $1$ of~\cite{CLT05}.

\begin{proposition} \hfil \label{prop_non_linear}
\begin{enumerate}
    \item \label{item_inn_product} For all $d, s, \theta\in \R$ and for all $u\in V_{d-\theta+s}$, $v\in V_{d-\theta-s}$, and $w\in w^{1+2\theta, \infty}$
        \begin{equation} \label{eq_nl_inn_product}
        \abs{ \ang{A^{d} B(u, v), w }_{1+2\theta} } \le C_{d, s, \theta} \nm{w}_{w^{1+2\theta, \infty}}
\abs{A^{(d-\theta+s)/2} u} \abs{A^{(d-\theta-s)/2} v},
        \end{equation}
where
\begin{equation} \label{eq_nl_const_inn_product}
C_{d, s, \theta} = \bigg ( \abs{a} (\lambda^{1 - 3d + 3\theta + s}
+ \lambda^{-(1 - 3d + 3\theta + s)}) + \abs{b} (\lambda^{2s} +
\lambda^{-(1 - 3d + 3\theta - s)}) \bigg ).
\end{equation}

    \item \label{item_inn_product2} For all $d, s, \theta\in \R$ and for all $w\in V_{d-\theta+s}$, $v\in V_{d-\theta-s}$, and $u\in w^{1+2\theta, \infty}$
        \begin{equation} \label{eq_nl_inn_product2}
        \abs{ \ang{A^{d} B(u, v), w }_{d-\theta+s} } \le c_{d, s, \theta} \nm{u}_{w^{1+2\theta, \infty}}
\abs{A^{(d-\theta+s)/2} w} \abs{A^{(d-\theta-s)/2} v},
        \end{equation}
where
\begin{equation} \label{eq_nl_const_inn_product2}
c_{d, s, \theta} = \bigg ( \abs{a} (\lambda^{- (2d - 2s)} +
\lambda^{2d - 2s}) + \abs{b} (\lambda^{1 + d - 3\theta - s} +
\lambda^{1 + d + 3\theta - s}) \bigg ).
\end{equation}

    \item \label{item_norm} For every $d, s\in \R$, the operator
$B : V_{2d-2s} \times V_{1+2s} \to V_{2d}$ and $B : V_{1+2s} \times
V_{2d-2s} \to V_{2d}$ is bounded and
        \begin{align} \label{eq_nl_norm}
        \babs{ A^{d} B(u, v) } \le \left\{%
\begin{array}{ll}
    c_{d, s - d, s} \nm{u}_{w^{1+2s, \infty}}
\abs{A^{d - s} v}, & \\
    C_{d, s - d, s} \nm{v}_{w^{1+2s, \infty}}
\abs{A^{d - s} u}, & \\
\end{array}%
\right.
        \end{align}
where the constant $C_{d, s, \theta}$ and $c_{d, s, \theta}$ were
defined in (\ref{eq_nl_const_inn_product}) and
(\ref{eq_nl_const_inn_product2}).

    \item \label{item_buvv} For every $u\in V_{d}$, $v\in V_{1-2d}$,
    for all $d\in \R$,
        \begin{equation} \label{eq_nl_buvv_eq}
        \ang{ B(u, v), u }_{d} = - \ang{ B(u, u), v }_{1-2d}^*,
        \end{equation}
and
        \begin{equation} \label{eq_nl_buvv}
        Re \ang{ B(v, u), u }_{d} = 0.
        \end{equation}


\end{enumerate}
\end{proposition}

\begin{proof}
To prove the inequality (\ref{item_inn_product}), we write
\begin{align*}
\babs{ \ang{A^{d} B(u, v), w }_{1+2\theta} } & = \babs{ \sum_{n 1}^\infty \bigg( a k_{n+1} k_n^{2d} v_{n+2} u_{n+1}^* w_{n}^* + b
k_n^{2d+1} v_{n+1} w_{n}^* u_{n-1}^* + \notag \\ & + a k_n^{2d}
k_{n-1} w_{n}^* u_{n-1} v_{n-2} + b k_n^{2d} k_{n-1} w_{n}^* v_{n-1}
u_{n-2} \bigg) } \le \notag
\\
& \le \sum_{n = 1}^\infty \babs{ a \lambda^{1 - 3d + 3\theta + s}
k_{n+2}^{d-\theta-s} v_{n+2} k_{n+1}^{d-\theta+s} u_{n+1}^*
k_n^{1+2 \theta} w_{n}^* } + \notag
\\
& + \babs{ b \lambda^{2s} k_{n+1}^{d-\theta-s} v_{n+1}
k_{n}^{1+2\theta} w_{n}^* k_{n-1}^{d-\theta+s} u_{n-1}^* } +
\notag
\\
& + \babs{ a \lambda^{-(1 - 3d + 3\theta + s)} k_n^{1+2\theta}
w_{n}^* k_{n-1}^{d-\theta+s} u_{n-1} k_{n-2}^{d-\theta-s} v_{n-2}
} + \notag
\\
& + \babs{ b \lambda^{-(1 - 3d + 3\theta - s)} k_n^{1+2\theta}
w_{n}^* k_{n-1}^{d-\theta-s} v_{n-1} k_{n-2}^{d-\theta+s} u_{n-2}
} \le \notag
\\
& \le C_{d, s, \theta} \nm{w}_{w^{1+2\theta, \infty}}
\abs{A^{(d-\theta+s)/2} u} \abs{A^{(d-\theta-s)/2} v}.
\end{align*}

In the same way we prove the inequality in
(\ref{item_inn_product2}).

In order to prove the statement (\ref{item_norm}) we apply
(\ref{eq_nl_inn_product2}) to obtain the first inequality
\begin{align*}
\babs{ A^{d} B(u, v) } & = \sup_{\abs{w} = 1} \babs{ ( A^{d} B(u,
v), w )} \le
\\
& \le \sup_{\abs{w} = 1} c_{d, s - d, s} \nm{u}_{w^{1+2s, \infty}}
\abs{w} \abs{A^{d-s} v} \le
\\
& \le c_{d, s - d, s} \nm{u}_{w^{1+2s, \infty}} \abs{A^{d-s} v}.
\end{align*}
The second inequality is proved similarly.

Finally, the statements (\ref{item_buvv}) follow directly from the
definition of the bilinear operator $B(u, v)$, the energy
conservation condition (\ref{eq_energy_cons_assumption}) and the
inequality (\ref{eq_nl_inn_product}).

\end{proof}

\section{Weak solutions of the inviscid shell model}

Let us consider the inviscid sabra shell model problem
\begin{subequations}
\label{eq_abstract_model_inviscid}
\begin{gather}
\pdt{u} + B(u, u) = f \label{eq_shell_model_inviscid} \\
u(0) = u^{in}. \label{eq_initial_cond_inviscid}
\end{gather}
\end{subequations}
One of the main properties of the sabra shell model of turbulence is
the locality of the non-linear interaction. This property allows us
to prove the global existence of weak solutions with the finite
energy to the inviscid shell model in the following sense. Similar
results in the context of the GOY shell model with the stochastic
forcing were obtained recently in (\cite{BBBF06_Stochastic}). In the
rest of the section we give the sufficient criteria for the weak
solutions to conserve the energy, and investigate the question of
the uniqueness of weak solutions.

\begin{definition} \label{def_weak}
Let $0 < T < \infty$, then $u(t)\in L^\infty([0, T], H) \bigcap
C([0, T], H_w)$ is called a weak solution of the system
(\ref{eq_abstract_model_inviscid}) on the interval $[0, T]$ if for
every $0 \le t \le T$ it satisfies
\begin{equation} \label{eq_weak_solution}
\ang{ u(t), v } + \int_0^t \ang{ B(u(s), u(s)), v } ds = \ang{
u^{in}, v } + \ang{ f, v },
\end{equation}
for every $v\in V$.
\end{definition}

Observe that if $u(t) = (u_1(t), u_2(t), u_3(t) \dots)$, then $u\in
C([0, T], H_w)$ is equivalent to $u_n(t)\in C([0, T], \C)$, for
every $n=1, 2, 3, \dots$.

\begin{theorem} \label{thm_weak}
Let $u^{in}, f\in H$, then for every $0 < T < \infty$ a weak
solution
\begin{equation} \label{eq_solution_reg}
u(t)\in L^\infty([0, T], H) \bigcap C([0, T], H_w),
\end{equation}
in the sense of Definition~\ref{def_weak} exists. In addition,
\begin{equation} \label{eq_solution_derivative_reg}
\frac{du(t)}{dt} \in L^\infty([0, T], V_{-1}).
\end{equation}
\end{theorem}

\begin{proof}
Let us fix $m > 1$. Denote by $P_m$ -- the orthogonal projection in
$H$ onto the first $m$ coordinates and $Q_m = I - P_m$. The Galerkin
approximating system of order $m$ for equation
(\ref{eq_abstract_model_inviscid}) is an $m$-dimensional system of
ordinary differential equations
\begin{subequations}
\label{eq_abstract_model_galerkin_inviscid}
\begin{gather}
\frac{du^m}{dt} + P_m B(u^m, u^m) = P_m f \label{eq_galerkin_inviscid} \\
u^m(0) = P_m u^{in}. \label{eq_galerkin_ic_inviscid}
\end{gather}
\end{subequations}
First observe that the nonlinear term of equations
(\ref{eq_abstract_model_galerkin_inviscid}) is quadratic in $u^m$.
Therefore, by the theory of ordinary differential equations, the
system (\ref{eq_abstract_model_galerkin_inviscid}) has a unique
solution on some finite time interval $[0, T^*_m)$. Let us now take
the inner product of both sides of the equation
(\ref{eq_galerkin_inviscid}) with $u_m$ and using subsequently
inequality (\ref{eq_nl_buvv}) of Proposition~\ref{prop_non_linear}
and the Cauchy-Schwartz inequality we get
\begin{equation} \label{eq_galerkin_energy_bound}
\half \ddt \abs{u^m}^2 = (P_m f, u^m) \le \abs{P_m f} \abs{u^m} \le
\abs{f} \abs{u^m},
\end{equation}
from which we conclude that
\begin{equation} \label{eq_galerkin_norm_bound}
\abs{u^m(t)} \le \abs{u^m(0)} + \abs{f} t \le \abs{u^{in}} + \abs{f}
t.
\end{equation}
Therefore, $u_m$ is finite in the $H$ norm for all $t < \infty$,
hence we can extend the solution of the problem
(\ref{eq_abstract_model_galerkin_inviscid}) to all $t\in [0,
\infty)$.

Let us fix $0 < T < \infty$. Then, from the relation
(\ref{eq_galerkin_norm_bound}) we may conclude that
\[
\sup_{0 \le t \le T} \abs{u_n^m(t)} \le C,
\]
for some constant $C > 0$ depending only on $u^{in}, f$, and $T$.
Moreover, writing the equation (\ref{eq_galerkin_inviscid}) in the
componentwise form
\begin{align}
u_n^m(t) & = u_n^m(0) + \int_0^t i \bigg( a k_{n+1} u^m_{n+2}
(u^m_{n+1})^* + b k_{n} u^m_{n+1} (u^m_{n-1})^* + \notag \\ & + a
k_{n-1} u^m_{n-1} u^m_{n-2} + b k_{n-1} u^m_{n-1} u^m_{n-2} \bigg)
ds + f_n. \label{eq_galerkin_inviscid_single_mode}
\end{align}
For $0 \le t \le T$, we get that for every $n$ there exists a
constant $C_n$, independent of $m$, such that
\[
\nm{u_n^m}_{C^1([0, T], \C)} \le C_n.
\]
Applying Arzela-Ascoli theorem we conclude that for every $n$ there
exists a subsequence $(m_k^n)_{k\ge 1}$ such that $u_n^{m_k^n}$
converges uniformly to some $u_n$, as $k\to \infty$. Moreover, by a
diagonalizing procedure we can choose a sequence $(m_k)_{k\ge 1}$,
independent of $n$ such that $u_n^m$ converges uniformly to $u_n \in
C([0, T], \C)$ and we denote
\[
u(t) = (u_1(t), u_2(t), u_3(t), \dots).
\]
Using the uniform convergence it is easy to show, passing to the
limit in the expression (\ref{eq_galerkin_inviscid_single_mode}),
that $u(t)$ satisfies the weak form of the sabra shell model
equation in the form
\begin{equation} \label{eq_weak_solution_weak}
\ang{ u(t), v^n } + \int_0^t \ang{ B(u(s), u(s)), v^n } ds = \ang{
u^{in}, v^n } + \ang{ f, v^n },
\end{equation}
for every $v^n\in H$ with the finite number of components different
from zero.

Moreover, we need to show that $u(t)\in L^\infty([0, T], H)$. The
sequence $\{u^m\}_{m\ge 1}$ is uniformly bounded in $H$ (see
(\ref{eq_galerkin_norm_bound})), and hence
\begin{equation} \label{eq_galerkin_bound_norm}
u^m \text{ is bounded in every } L^p([0, T], H), \text{ for } 1 \le
p\le \infty.
\end{equation}
Therefore, we conclude that there exists a subsequence $(m_k)_{k\ge
1}$ such that $u^{m_k}$ converges to $w(t)$ in the weak-* topology
of $L^\infty([0, T], H)$, and by definition it is not hard to see
that the limiting function is indeed $w(t) \equiv u(t)$. In
addition, by the inequality (\ref{item_norm}) of
Proposition~\ref{prop_non_linear}, we get
\[
\abs{B(u, u)}_{-1} = \abs{A^{-1/2} B(u, u)} \le C \abs{u}^2,
\]
for $C = C_{-1/2, 0, -1/2}$ (see (\ref{eq_nl_const_inn_product})),
concluding that $B(u, u)\in L^\infty([0, T], V_{-1})$.

Finally, let $v\in V$, and $v^n = P_n v$, with the finite number of
components being different from zero, converging strongly to $v$.
Then letting $n\to \infty$ in the relation
(\ref{eq_weak_solution_weak}) we conclude that $u(t)$ satisfies the
equation (\ref{eq_abstract_model_inviscid}) in the weak sense of
Definition~\ref{def_weak}.

\end{proof}

The next Theorem gives a partial answer to the question: under which
conditions the weak solutions conserve the energy?

\begin{theorem} \label{thm_weak_smooth}
Let $u(t)$ be a weak solution, whose existence is proved in
Theorem~\ref{thm_weak}, satisfying
\begin{equation} \label{eq_solution_more_reg}
u(t)\in L^\infty([0, T], V_{1/3}),
\end{equation}
for some $T > 0$. Then for every $t\in [0, T]$
\begin{equation} \label{eq_conservation_energy}
\abs{u(t)}^2 = \abs{u^{in}}^2 + \int_0^t (f, u(s)) ds.
\end{equation}
\end{theorem}

\begin{proof}
If a weak solution satisfies (\ref{eq_solution_more_reg}), then
according to the inequality (\ref{item_buvv}) of
Proposition~\ref{prop_non_linear}, the equation
(\ref{eq_abstract_model_inviscid}) can be considered as the equation
in the space $V_{-1/3}$. Applying the operator $A^{-1/3}$ to both
sides of equation (\ref{eq_abstract_model_inviscid}), and taking an
inner product with $A^{1/3} u(t)$ in the space $H$ we get, using the
identity (\ref{item_buvv}) of Proposition~\ref{prop_non_linear},
\[
\half \ddt \abs{u(t)}^2 = (f, u(t)),
\]
from which the statement follows.
\end{proof}

As we already mentioned in the introduction, our result for the
sabra shell model of turbulence is reminiscent of the Onsager's
conjecture for the Euler equations (see \cite{CET94_Onsager},
\cite{DuchRob00_Onsager}, \cite{Eyink94}, \cite{Onsa49_Hydro}).
However, the criterion given by Theorem~\ref{thm_weak_smooth} is not
sharp. It is easy to give an example of a solution of the inviscid
sabra shell model of turbulence, which stays merely in $H$, but
still conserves the energy. To see this, consider the forcing $f (f_1, f_2, \dots)$, where
\[
f_n = \left\{%
\begin{array}{ll}
    \frac{1}{n}, & n = 1, 3, 6, \dots, \\
    0, & \text{o/w}. \\
\end{array}%
\right.
\]
Then solution $u(t) = (u_1(t), u_2(t), \dots)$, where $u_n(t) \frac{t}{n}$, for $n = 1, 3, 6, \dots$, is a weak solution of the
sabra shell model, corresponding to the zero initial condition. The
function $u(t)$ is only in $H$ for every $t < \infty$, however it is
easy to see that it conserves the energy. Clearly, this example is
pathological in a sense that all non-linear interactions are absent
due to the wide gaps between the excited modes, however it shows
that the result of Theorem~\ref{thm_weak_smooth} is not sharp.
Moreover, it is not known when the weak solutions of the inviscid
sabra shell model dissipate energy. These questions will be studied
in the forthcoming work \cite{CLT06_Dissipation}.

The final result of this section gives criterions for the uniqueness
of weak solutions.

\begin{theorem} \hfil \label{thm_weak_unique}
\begin{enumerate}
\item \label{thm_weak_unique_1} Let $u(t), v(t)$ be two weak solutions, whose
 existence is proved in
Theorem~\ref{thm_weak}, satisfying
\begin{equation} \label{eq_solution_unique}
u(t), v(t) \in L^1([0, T], w^{1, \infty}),
\end{equation}
for some $T > 0$, and $u(0) = v(0)$. Then $u(t) = v(t)$, for all
$t\in [0, T]$.

\item \label{thm_weak_unique_2} If $u(t), v(t)$ are two weak solutions,
satisfying
\begin{equation} \label{eq_solution_unique_21}
u(t)\in L^1([0, T], w^{1, \infty}) \bigcap L^\infty([0, T],
V_{1/3}),
\end{equation}
and
\begin{equation} \label{eq_solution_unique_22}
v(t)\in L^\infty([0, T], V_{1/3}),
\end{equation}
for some $T > 0$, with $u(0) = v(0)$. Then $u(t) = v(t)$, for all
$t\in [0, T]$.

\end{enumerate}
\end{theorem}

\begin{proof}
First, let $u, v\in L^1([0, T], w^{1, \infty})$ be two weak
solutions of the inviscid sabra shell model with the same initial
conditions. Denote $w = u - v$ satisfying
\[
\frac{dw}{dt} + B(u, w) + B(w, v) = 0,
\]
with $w(0) = 0$. Using the fact that $u, v$ and $w$ satisfy
(\ref{eq_solution_derivative_reg}) and (\ref{eq_solution_reg}), we
are allowed to apply the operator $A^{-1/2}$ to both sides of the
last equation and then take the inner product of both sides with $w$
in $H$ to conclude
\begin{align} \label{eq_uniq_ineq}
\half \ddt \abs{A^{-1/4} w}^2 & \le \abs{ (A^{-1/2} B(u, w), w) } +
\abs{ (A^{-1/2} B(w, v), w) } = \notag
\\
& = \abs{ (A^{-1/2} B(u, w), w) } + \abs{ (B(w, A^{-1/2} w), v) }
\le \notag
\\
& \le C_1 (\nm{v}_{w^{1, \infty}} + \nm{u}_{w^{1, \infty}})
\abs{A^{-1/4} w}^2,
\end{align}
where we subsequently used parts (\ref{item_buvv}),
(\ref{item_inn_product}), (\ref{item_inn_product2}) of
Proposition~\ref{prop_non_linear} and \newline $C_1 = C_{-1/2, -1/2,
0} + c_{-1/2, 0, 0}$. Applying the Gronwall's inequality to
(\ref{eq_uniq_ineq}) we get
\[
\abs{A^{-1/4} w(t)}^2 \le \abs{A^{-1/4} w(0)}^2 e^{C_1 \int_0^t
(\nm{v(s)}_{w^{1, \infty}} + \nm{u(s)}_{w^{1, \infty}}) ds},
\]
for $t\in [0, T]$, concluding the proof of the part
(\ref{thm_weak_unique_1}).

To prove the part (\ref{thm_weak_unique_2}) of the theorem, let
$u(t)$ be the solution of the inviscid sabra shell model satisfying
(\ref{eq_solution_unique_21}). Let $v(t)$, satisfying
(\ref{eq_solution_unique_22}), be another weak solution with the
same initial data $v(0) = u(0)$. Note that, in particular, both
$u(t)$ and $v(t)$ conserve the energy, according to
Theorem~\ref{thm_weak_smooth}. Denote $w = u - v$ satisfying
\begin{equation} \label{eq_w_eq}
\frac{dw}{dt} + B(u, w) + B(w, u) + B(w, w) = 0,
\end{equation}
with $w(0) = 0$. Just as in the proof of
Theorem~\ref{thm_weak_smooth}, we can consider (\ref{eq_w_eq}) as an
equation in the space $V_{-1/3}$. Therefore, applying the operator
$A^{-1/3}$ to both sides of equation
(\ref{eq_abstract_model_inviscid}), and taking an inner product with
$A^{1/3} w(t)$ in the space $H$ we get, using the parts
(\ref{item_inn_product}) and (\ref{item_buvv}) of
Proposition~\ref{prop_non_linear},
\[
\half \ddt \abs{w}^2 \le \abs{\ang{B(w, u), w}_{1/3}} \le C_2
\nm{u}_{w^{1, \infty}} \abs{w}^2,
\]
for $C_2 = C_{0, 0, 0}$. It follows that
\[
\abs{w(t)}^2 \le \abs{w(0)}^2 e^{C_2 \int_0^t \nm{u(s)}_{w^{1,
\infty}} ds},
\]
for $t\in [0, T]$, finishing the proof of the theorem.
\end{proof}

\section{The short-time existence and uniqueness of strong solutions}

The uniqueness of the weak solutions, whose existence was proved in
Theorem~\ref{thm_weak} is not known. In this section we prove that
the weak solutions of the inviscid case sabra shell model are unique
as long as they stay smooth enough, at least on the short-time
interval $[0, T_*)$, where the time $T_*$ depends on the parameters
of the problem ($a, b, c, k_0$ and $\lambda$), as well as the
initial data $u^{in}$ and $f$. Let us consider the inviscid sabra
shell model problem (\ref{eq_abstract_model_inviscid}) as an
ordinary differential equation (ODE) in the Hilbert space $V_{d}$,
for $d \ge 1$. The main theorem of this section shows the short time
existence and uniqueness of solutions of the equation
(\ref{eq_abstract_model_inviscid}).

\renewcommand{\labelenumi}{(\roman{enumi})}
\begin{theorem} \label{thm_existence}
Let $u^{in}\in V_d$ and $f\in V_{d}$ for some $d \ge 1$.
\begin{enumerate}
    \item \label{thm_existence_ex} There exists a time $T > 0$, such that the inviscid problem
(\ref{eq_abstract_model_inviscid}) has a unique solution $u(t)$
satisfying
\[
u(t)\in C^1((-T, T), V_d).
\]

    \item \label{thm_existence_reg} Moreover, if $f\in V_{2d-1}$, then
\[
\frac{du}{dt}\in C((-T, T), V_{2d-1}).
\]

    \item \label{thm_existence_cont} The unique solution to the inviscid sabra shell model (\ref{eq_abstract_model_inviscid}) either exists globally in time, or there exists a maximal positive time of existence $T_* > 0$
such that
\[
u(t)\in C^1([0, T_*), V_d),
\]
and
\[
\limsup_{t\to T_*^-} \abs{u(t)}_d = \infty.
\]
A similar statement can be formulated for the maximal negative time
of existence.

\end{enumerate}
\end{theorem}

In our notation $C^1((-T, T), V_d)$ denotes continuously
differentiable functions on the interval $(-T, T)$ with values in
$V_d$. Our proof is based on the classical Picard theorem for ODEs
in Banach spaces (see, for example, \cite{Har82}, \cite{MB02},
\cite{Schechter04}).

\begin{proof} (Of Theorem~\ref{thm_existence})

Let us write the system (\ref{eq_abstract_model_inviscid}) in the
form
\begin{equation} \label{eq_abstract_model_inviscid_ode}
\pdt{u} = F(u), \;\;\; u(0) = u^{in},
\end{equation}
where $F(u) = f - B(u, u)$. Fix $d\ge 1$, then according to the part
(\ref{item_norm}) of Proposition~\ref{prop_non_linear} the operator
$B(u, u)$ maps $V_d$ into $V_{2d - 1}\subseteq V_d$. Hence, the
mapping $F(u)$ maps $V_d$ into itself. Moreover, for every $u, v\in
V_d$ we have the following estimates
\begin{align*}
\abs{F(u) - F(v)}_d & = \abs{B(u, u) - B(v, v)}_d \le
\\
& \abs{B(u - v, u)}_d + \abs{B(v, u - v)}_d \le C_2 (\nm{u} +
\nm{v}) \abs{u - v}_d,
\end{align*}
where the last inequality follows from relation (\ref{eq_nl_norm}),
and $C_2 = C_{d, -d, 0} + c_{d, -d, 0} > 0$ (see
(\ref{eq_nl_const_inn_product})). Therefore, we conclude that the
mapping $F(u)$ is locally Lipschitz continuous, and we are able to
apply the Picard theorem, completing the proof.

Part (\ref{thm_existence_cont}) follows by the straightforward
application of classical theory of ODEs. To prove part
(\ref{thm_existence_reg}) we apply inequality (\ref{eq_nl_norm}) to
both sides of the equation (\ref{eq_abstract_model_inviscid}).

\end{proof}

Using the part (\ref{thm_existence_cont}) of
Theorem~\ref{thm_existence} we will be able in the
section~\ref{sec_bkm} to derive a criterion for the blow-up of the
solutions of the inviscid sabra shell model and to prove the global
(in time) existence of the unique, regular solutions in the
particular case of the $2$-D regime of the inviscid sabra shell
model.

\section{A Beale-Kato-Majda type result} \label{sec_bkm}

The Beale-Kato-Majda theorem (see \cite{BKM84}, \cite{MB02}) states,
citing the original article, ``if a solution of the Euler or
Navier-Stokes equations is initially smooth and loses its regularity
at some later time, then the maximum vorticity necessarily grows
without bound as the critical time approaches.'' More precisely, if
the initially smooth solution of the Euler equations cannot be
continued beyond the time $T^*$, and $T^*$ is the first such time,
then
\[
\lim_{t\to {T^*}^-} \int_0^{t} \nm{\omega(\cdot, s)}_{L^\infty} ds
= \infty,
\]
where $\omega = curl v$ is the vorticity and $v$ is the velocity
field of the Euler equations.

Our goal in this section is to derive a similar criterion for the
loss of regularity of the solutions of the inviscid sabra shell
model. Note that in our case, the analog of the $L^\infty$ norm of
the vorticity would be the $\ell^{\infty}$ norm of the velocity
``derivative'', namely
\[
\nm{u}_{w^{1, \infty}} = \sup_{1\le n \le \infty} k_n \abs{u_n}.
\]
Clearly, if this quantity becomes infinite at some finite moment of
time, then all higher norms, namely $\abs{u}_d$, for $d\ge 1$,
become unbounded at the same time. However, in the spirit of the
Beale-Kato-Majda result for the Euler equations, we show that the
opposite is also true. In other words, we show that if a regular
solution $u(t)$ of the inviscid shell model problem loses its
regularity for the first time at the time $T$, then
\[
\int_0^{t} \nm{u(s)}_{w^{1, \infty}} ds \longrightarrow \infty,
\;\;\; \text{as} \;\;\; t\to {T}^-.
\]

For simplicity we would like to focus on  the inviscid sabra shell
model problem (\ref{eq_abstract_model_inviscid}) without forcing
\begin{subequations}
\label{eq_abstract_model_inviscid_no_force}
\begin{gather}
\pdt{u} + B(u, u) = 0 \label{eq_shell_model_inviscid_no_force} \\
u(0) = u^{in}(x). \label{eq_initial_cond_inviscid_no_force}
\end{gather}
\end{subequations}

\begin{theorem} \label{thm_bkm}
Let $u^{in}\in V_d$, for some $d \ge 1$. Let $u(t)\in C^1([0, T_*),
V_{d})$ be the solution of the inviscid shell model equation
(\ref{eq_abstract_model_inviscid_no_force}), where $T_*$ is its
maximal positive time of existence. Then, either $T_* = \infty$ or
\[
\lim_{t\to T_*^-} \int_0^{t} \nm{u(\tau)}_{w^{1, \infty}} d\tau \infty,
\]
and hence
\[
\limsup_{t\to T_*^-} \nm{u(t)}_{w^{1, \infty}} = \infty.
\]
\end{theorem}

\begin{proof}

Let us fix $d \ge 1$ and consider $u(t)$ -- the unique solution to
the sabra shell model equation
(\ref{eq_abstract_model_inviscid_no_force}). According to the part
(\ref{thm_existence_reg}) of Theorem~\ref{thm_existence}, both sides
of the equation (\ref{eq_shell_model_inviscid_no_force}) lie in the
space $V_{2d-1}$. Therefore, we are allowed to apply the operator
$A^{d/2}$ to both sides of the equation and take the inner product
in $H$ with $A^{d/2} u \in H$. After taking the real part we obtain
\begin{align} \label{eq_nl_equality}
\half \ddt \abs{u}_d^2 & = Re (A^{d/2} B(u, u), A^{d/2} u) = \notag
\\
& = - Im \sum_{n = 1}^\infty \bigg( a k_{n+1} k_n^{2d} u_{n+2}
u_{n+1}^* u_{n}^* + b k_n^{2d+1} u_{n+1} u_{n}^* u_{n-1}^* + \notag
\\
& + a k_n^{2d} k_{n-1} u_{n}^* u_{n-1} u_{n-2} + b k_n^{2d} k_{n-1}
u_{n}^* u_{n-1} u_{n-2} \bigg) = \notag
\\
& = E_d Im \sum_{n = 2}^\infty k_n^{2d + 1} u_{n+1} u_n^* u_{n-1}^*,
\end{align}
where we denote by
\begin{equation} \label{eq_nl_constant_ed}
E_d = a (\lambda^{2d} - \lambda^{-2d}) + b (\lambda^{2d} - 1).
\end{equation}
Applying Cauchy-Schwarz inequality to (\ref{eq_nl_equality}) we get
\[
\ddt \abs{u}_d^2 \le 2 \abs{E_d} \nm{u}_{w^{1, \infty}} \abs{u}_d^2,
\]
which is valid for $t\in [0, T]$, and for every $T < T_*$. Using
Gronwall's inequality we conclude
\begin{equation} \label{eq_apriori_bound_333}
\abs{u(t)}_d \le \abs{u(0)}_d e^{\abs{E_d} \int_0^T
\nm{u(\tau)}_{w^{1, \infty}} d\tau}.
\end{equation}
The Theorem follows after letting $T\to T_*^-$ in the last
inequality.

\end{proof}

\section{The ``two-dimensional'' regime} \label{sec_2d}

Recall that for the range of parameters satisfying $0 < c/a < 1$,
corresponding to the ``two-dimensional'' regime, the sabra shell
model possesses two different positive quadratic invariants, one is
associated with the energy and the second is associated with the
enstrophy in the analogy with the $2$-D Euler equation (see
\cite{GLP02})
\begin{equation} \label{eq_enstrophy_norm}
\mathrm{W} = \abs{A^{d_{0}/2} u} = \abs{u}_{d_{0}},
\end{equation}
where
\begin{equation} \label{eq_enstrophy_norm_exp}
d_{0} = \half \log_\lambda \bigg( \frac{a}{c} \bigg).
\end{equation}
In this case $E_{d_{0}}$, defined in (\ref{eq_nl_constant_ed}),
equals $0$. Moreover, if $d_{0} \ge 1$, the following inequality
holds
\[
\nm{u}_{w^{1,\infty}} \le \abs{u}_{d_{0}},
\]
for every $u\in V_{d_{0}}$. From the relation
(\ref{eq_enstrophy_norm_exp}) we conclude that the condition $d_0
\ge 1$ corresponds to the case when parameters of the inviscid sabra
shell model satisfy
\begin{equation}\label{eq_par_global}
0 < \frac{c}{a} \le \lambda^{-2}.
\end{equation}

It is well known that the Euler equations of the ideal
incompressible fluid in $2$-D possesses global in time, unique,
regular solution (see, for example, \cite{Bard72_Euler},
\cite{Bard72_Euler2}, \cite{Kato}, \cite{MB02}, \cite{MP94},
\cite{Yu63}). The same statement is true for the inviscid shell
model of turbulence, namely.

\begin{corollary} \label{cor_2d_global} (Global Existence)
Let $d_0$ be defined by the relation (\ref{eq_enstrophy_norm_exp}).
\begin{enumerate}
\item \label{cor_2d_global_weak} Let the parameters $a, c$, and $\lambda$ of the inviscid sabra shell
model (\ref{eq_abstract_model_inviscid_no_force}) satisfy
\[
\frac{c}{a} > \lambda^{-2}.
\]
Then for $u^{in}\in V_{d_0}$, there exists a weak solution $u(t)$ to
the inviscid problem (\ref{eq_abstract_model_inviscid_no_force})
satisfying
\[
u(t)\in L^\infty((-\infty, \infty), V_{d_0}).
\]

\item \label{cor_2d_global_weak_onsager} The weak solution $u(t)$ conserves the enstrophy (\ref{eq_enstrophy_norm}), for all $t\in [0, T]$, provided
\[
u(t)\in L^\infty([0, T], V_{(1 - 2d_0)/3}).
\]

\item \label{cor_2d_global_strong} If the parameters $a, c$, and $\lambda$ of the inviscid sabra shell
model (\ref{eq_abstract_model_inviscid_no_force}) satisfy the
relation (\ref{eq_par_global}), then for $u^{in}\in V_s$, $s \ge
d_0$, there exists a unique global solution $u(t)$ to the inviscid
problem (\ref{eq_abstract_model_inviscid_no_force}) satisfying
\[
u(t)\in C^1((-\infty, \infty), V_{s}).
\]
\end{enumerate}
\end{corollary}

The proof of the part (\ref{cor_2d_global_weak}) of
Corollary~\ref{cor_2d_global} is essentially the same as the proof
of Theorem~\ref{thm_weak}. The proof of the part
(\ref{cor_2d_global_weak}) is similar to that of
Theorem~\ref{thm_weak_smooth}, and the part
(\ref{cor_2d_global_strong}) follows from the criterion, proved in
Theorem~\ref{thm_bkm}.

The comprehensive numerical study of the shell model of turbulence
in the $2$-D parameters regime was performed in \cite{GLP02} for the
sabra model, and previously in \cite{DiMo96} for the GOY model. In
particular, it showed that parameters setting defined by relation
(\ref{eq_par_global}) corresponds to the enstrophy and energy
equipartition across the inertial range. Therefore, our rigorous
result on existence of the solutions of the inviscid shell model
(\ref{eq_abstract_model_inviscid_no_force}) globally in time
supports these findings.

It was also found numerically, that for the parameters satisfying
$\frac{c}{a}
> \lambda^{-2}$, the shell models exhibit the direct enstrophy
cascade in the inertial range and the energy distribution becomes
close to the Kraichnan's dimensional prediction
\[
\ang{\abs{u_n}^2} \sim k_n^{-\frac{2}{3} (1 + \log_\lambda(a/c))},
\]
with small corrections, for $n_f \ll n \ll n_d$, where $n_f$ is the
largest wavenumber of the forcing and $n_d$ is the Kraichnan's
dissipation wavenumber (see \cite{DiMo96}). Note that for
$\frac{c}{a} = \lambda^{-2}$ this estimate exactly coincides with
the well-known prediction $k_n^{-3}$ for the energy spectrum of the
$2$-D developed turbulence (see \cite{Kraich67}). Using these
reasonings we conclude that for
\begin{equation} \label{eq_blowup_param_2d}
\frac{c}{a} > \lambda^{-2},
\end{equation}
the inertial range of the sabra shell model with non-zero viscosity
will scale like $\ang{\abs{u_n}^2} \sim k_n^{-2 + \delta}$, for some
positive $\delta$. Therefore, it is natural to expect that if the
viscosity tends to zero, or equivalently, the dissipation scale
$n_d$ tends to infinity, the solutions of the inviscid sabra shell
model, for parameters satisfying relation
(\ref{eq_blowup_param_2d}), will blow-up in finite time, for some
initial conditions, according to the criterion proved in
Theorem~\ref{thm_bkm}.

\section{Cubic invariant and Hamiltonian structure}

In practical numerical simulations of the sabra shell model one is
limited to consider a truncated model of $N$ equations, setting
$u_{n} = 0$, for $n = N+1, N+2, \dots$. It was shown in \cite{LPP99}
that such a finite system in the inviscid and unforced case
possesses a Hamiltonian structure for a specific value of the
parameters. In this section we will state, based on our results on
the existence of the solutions of the inviscid sabra shell model,
that the infinite system of equations also has a Hamiltonian
structure.

By rescaling the time and taking into the account the energy
conservation assumption (\ref{eq_energy_cons_assumption}), we will
assume that
\[
a = 1, \;\; b = -\epsilon, \;\; c = \epsilon -1.
\]
Let us fix $\epsilon = (\sqrt{5} - 1)/2$ -- the golden mean
satisfying $\epsilon^2 = 1 - \epsilon$. In that case we can rewrite
the equations (\ref{eq_sabra}) in the equivalent form
\begin{equation} \label{eq_sabra_eq_new}
\frac{d u_n}{d t} = i k_{n+1} (u_{n+2} u_{n+1}^* -
\frac{\epsilon}{\lambda} u_{n+1} u_{n-1}^* +
\frac{\epsilon^2}{\lambda^2} u_{n-1} u_{n-2}) - \nu k_n^2 u_n + f_n,
\end{equation}
for $n = 1, 2, 3, \dots$. In that case the inviscid sabra shell
model without forcing has, fomrally, a cubic invariant of the form
\begin{equation} \label{eq_cubic_invariant}
I = \sum_{n=1}^\infty \epsilon k_0 \bigg ( -
\frac{\lambda}{\epsilon} \bigg )^{n} (u_{n+1}^* u_n u_{n-1} + c.c.),
\end{equation}
where c.c. stands for complex conjugate.

Following the method of \cite{LPP99}, we perform the following
change of variables
\[
a_n = \frac{u_n}{\epsilon^{n/2}}, \;\;\; \text{for even} \; n,
\]
and
\[
a_n = - \frac{u_n^*}{\epsilon^{n/2}}, \;\;\; \text{for odd} \; n.
\]
The modified equations then take the form of
\begin{equation} \label{eq_ham_motion}
\pdt{a_n} = -i k_0 \epsilon (\lambda \sqrt{\epsilon})^n \bigg(
\lambda \sqrt{\epsilon} a_{n+2} a_{n+1} + a^*_{n+1} a_{n-1} +
\frac{1}{\lambda \sqrt{\epsilon}} a^*_{n-1} a_{n-2} \bigg),
\end{equation}
for $n = 1, 2, 3, ...$. Finally, the Hamiltonian takes the form
\begin{equation} \label{eq_hamiltonian}
\mathcal{H} = \sum_{n=1}^\infty \mathcal{H}_n,
\end{equation}
where
\begin{equation} \label{eq_hamiltonian_one}
\mathcal{H}_n = \epsilon k_0 (\lambda \sqrt{\epsilon})^n (a_{n+1}
a_{n} a_{n-1}^* + c.c.).
\end{equation}
In order to see that $\mathcal{H}$ is indeed a Hamiltonian we note
that
\[
\frac{d \mathcal{H}}{dt} = 0,
\]
and the equations of motion (\ref{eq_ham_motion}) satisfy
\[
\frac{d a_n}{d t} = - i \frac{\partial \mathcal{H}}{\partial
a_n^*}, \;\;\; \frac{d a^*_n}{d t} = i \frac{\partial
\mathcal{H}}{\partial a_n}.
\]

Both the cubic invariant $I$ and the Hamiltonian $\mathcal{H}$ are
defined by infinite sums. Therefore, it is natural to ask when those
definitions make sense, namely when the sums converge.

\begin{lemma} \label{lem_hamiltonian}
For $\lambda^2 \ge \epsilon^{-1}$ the Hamiltonian
(\ref{eq_hamiltonian}) is well defined for all $u\in V$.
\end{lemma}

\begin{proof}
The Lemma follows by a simple application of H\"{o}lder
inequality.
\end{proof}

Finally, we can conclude the following.

\begin{corollary}
Let the parameters of the inviscid sabra shell model
(\ref{eq_abstract_model_inviscid_no_force}) satisfy
\[
a = 1, \;\;\; b = -\frac{\sqrt{5} - 1}{2}, \;\;\; c = \frac{\sqrt{5}
- 3}{2}, \;\;\; \lambda^2 \ge \frac{2}{\sqrt{5} - 1}.
\]
Then the inviscid sabra shell model
(\ref{eq_abstract_model_inviscid_no_force}) with initial data in
$V_d$, for $d\ge 1$, is a Hamiltonian system with the Hamiltonian
defined by relation (\ref{eq_hamiltonian}), as long as a solution of
the model exists.
\end{corollary}

\section{Conclusions}

In this work we continued the analytic study of the shell models of
turbulence, initiated in~\cite{CLT05}. We established the global
existence of weak solutions and showed that strong solutions remain
regular and unique for some short period of time. Moreover, we
showed that the solutions for the ``two-dimensional'' range of
parameters remain regular and unique globally in time. In addition,
we established a Beale-Kato-Majda type criterion for the blow-up of
the initially smooth solutions.

We showed that for some parameter regime the sabra shell model is an
infinite dimensional Hamiltonian system. In contrast to the Euler
equations, which possess a quadratic Hamiltonian function (see, for
example, \cite{ArnKhe98}, \cite{MaRa99}), the Hamiltonian of the
inviscid sabra shell model is cubic.

We showed that the weak solution $u(t) = (u_1(t), u_2(t), \dots)$
conserve the energy provided that the components of the solution
satisfy
\[
\abs{u_n} \le C k_n^{-1/3} (\sqrt{n} \log(n+1))^{-1},
\]
for some positive absolute constant $C$. A similar result for the
Euler equations is known as the Onsager's conjecture (see
\cite{Onsa49_Hydro}) and it was proved in \cite{CET94_Onsager} (see
also \cite{DuchRob00_Onsager}, \cite{Eyink94}). The question of
whether less regular solutions dissipate energy remains open.

The question of the possible loss of regularity and uniqueness for
the initially smooth solutions of the inviscid sabra shell model
outside of the ``two-dimensional'' range of parameters still remains
open.

The dimensional argument for the viscous ($\nu > 0$) GOY shell
model, which are also applicable to the sabra model, indicate that
in the ``three dimensional'' parameters regime $-1 < \frac{c}{a} <
0$ the velocity field scales like
\[
\ang{\abs{u_n}} \sim k_n^{-\frac{1}{3} (1 + \log_\lambda\abs{a/c})},
\]
at the inertial range (see \cite{DiMo96}). Therefore, at least for
the parameters regime satisfying
\begin{equation} \label{eq_blowup_param_3d}
\babs{\frac{c}{a}} > \lambda^{-2},
\end{equation}
we might expect the blow-up of the inviscid sabra (as well as GOY)
shell model of turbulence.

We would like to mention that the techniques, used to prove the
blow-up for other discrete models of Euler equations (see, e.g.,
\cite{Chesk06}, \cite{FriPav04}, \cite{KaPav05},
\cite{KiZla05_Blowup}, \cite{Wal05}) could not be applied directly
in the case of the sabra shell model of turbulence. The study of
possible loss of regularity of the inviscid sabra shell model in
different parameters regime is the subject of ongoing work
(\cite{CLT06_Dissipation}).

\noindent
\section*{Acknowledgments}

The authors would like to thank I.~Procaccia, V.~Lvov and
A.~Pomyalov for the very stimulating and inspiring discussions.
E.S.T. is thankful to the kind hospitality of the \'{E}cole Normale
Sup\'{e}rieure - Paris where this work was completed. The work of
P.C.~was partially supported by the NSF grant No.~DMS-0504213. The
work of E.S.T. was supported in part by the NSF grant
No.~DMS--0504619, the MAOF Fellowship of the Israeli Council of
Higher Education, and by the BSF grant No.~2004271.

\end{document}